\documentclass[aps,preprint,amsmath,amssymb,superscriptaddress]{revtex4}

\setcitestyle{super}

\usepackage{setspace}
\usepackage{graphicx}% Include figure files
\usepackage{dcolumn}% Align table columns on decimal point
\usepackage{bm}% bold math
\usepackage{tabularx}
\usepackage{multirow}
\newcommand{\nc}{\newcommand}
\nc{\be}{\begin{equation}}
\nc{\ee}{\end{equation}}
\nc{\bea}{\begin{eqnarray}}
\nc{\eea}{\end{eqnarray}}
\nc{\bean}{\begin{eqnarray*}}
\nc{\eean}{\end{eqnarray*}}
\nc{\mb}{\mbox}
\nc{\rnc}{\renewcommand}
\nc{\vk}{\mb{\bf k}}
\nc{\vp}{\mb{\bf p}}
\nc{\vn}{\mb{\bf n}}
\nc{\vq}{\mb{\bf q}}
\nc{\rr}{\mb{\bf r}}
\nc{\vz}{\hat {\mb{\bf z}}}
\nc{\vj}{\mb{\boldmath$j$}}
\nc{\vg}{\mb{\boldmath$g$}}
\nc{\x}{\mb{\boldmath$x$}}
\nc{\A}{\mb{\boldmath$A$}}
\nc{\va}{\mb{\boldmath$a$}}
\nc{\vs}{\mb{\boldmath$\sigma$}}
\nc{\vpi}{\mb{\boldmath$\pi$}}
\nc{\nab}{\nabla}
\nc{\X}{\sf x}
\nc{\asp}{\hspace{10mm}}

\begin{document}
\begin{spacing}{2}

\title{Theory of $t_{2g}$ electron-gas Rashba interactions}

\author{Guru Khalsa}
\email[email:]{guru@physics.utexas.edu,macd@physics.utexas.edu}
\affiliation{Department of Physics, University of Texas at Austin, Austin TX 78712, USA}

\author{Byounghak Lee}
\affiliation{Department of Physics, Texas State University, San Marcos, TX 78666, USA}

\author{A.H. MacDonald}
\email[email:]{guru@physics.utexas.edu,macd@physics.utexas.edu}
\affiliation{Department of Physics, University of Texas at Austin, Austin TX 78712, USA}

\maketitle

{\bf The spin-degeneracy of Bloch bands in a crystal can be lifted\cite{Kittel,Dresselhaus} when spin-orbit (SO) coupling is 
present and inversion symmetry is absent.  In two-dimensional electron systems (2DES)  
spin-degeneracy is lifted by Rashba interaction\cite{Rashba,Winkler} terms -
symmetry invariants that are scalar products of spin and orbital axial vectors.
Rashba interactions are symmetry allowed whenever a 2DES is not invariant under reflections through the plane it occupies.   
In this paper, we use a tight-binding model informed by {\it ab initio} electronic structure 
calculations to develop a theory of Rashba splitting in the $t_{2g}$ bands of the two-dimensional 
electron systems\cite{Hwang,Mannhart,Ueno2008,Ueno2011,OxideInterfaces}
formed at cubic perovskite crystal surfaces and interfaces.  We find that 
Rashba splitting in these systems is due to atomic-like on-site 
SO interactions combined with processes in which $t_{2g}$ electrons change orbital character when they hop between 
metal sites.  These processes are absent in a cubic 
environment and are due primarily to polar lattice distortions which alter the metal-oxygen-metal bond angle.}

\begin{figure}[H]
\includegraphics[width=8.6cm]{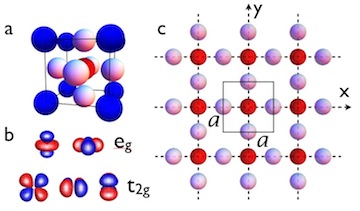} 
\caption{{\bf Perovskite crystal structure. a}, Bulk cubic unit cell with the A atom in blue, B atom in red, and the oxygen in white. {\bf b,} Splitting of atomic d-orbitals into $e_g$ and $t_{2g}$ manifolds.  {\bf c,} Single BO$_2$ plane with one unit cell, with area $a^2$, in the boxed region.  
} 
\label{CrystalStructure}
\end{figure}

Bulk cubic perovskites have chemical formula ABO$_3$     
and the crystal structure illustrated in Figure \ref{CrystalStructure}a. 
The 2DESs in which we are interested are formed  
from conduction band $B$-site transition metal $d$-orbitals.
Because the B site, at the cubic cell center, has octahedral coordination\cite{Goodenough} with neighboring oxygen atoms,
located at the centers of the cubic cell faces, oxygen-metal bonding partially  
lifts the degeneracy of the d-orbitals, pushing the $e_g = \{x^2-y^2,3z^2-r^2\}$ orbitals up in energy relative to
the $t_{2g} = \{yz,zx,xy\}$ orbitals (Figure \ref{CrystalStructure}b).
In the the simplest model of the bulk electronic structure\cite{Mattheiss}, the bonding networks of the 
three $t_{2g}$ bands are decoupled; an $xy$-orbital on one B site, for example,  
can hop only along the $y$ or $x$ direction through an
intermediate $p_x$  or $p_y$ orbital to an $xy$-orbital on the B site of a neighboring cubic cell. 
In perovskite 2DESs, the $t_{2g}$ bands are reconstructed\cite{Khalsa,Stengel} into 2D subbands
whose detailed form depends on the bulk band parameters\cite{Bistritzer}, the surface or interface confinement mechanism,   
and the dielectric response of the material.  
A polar displacement
of A and B atoms relative to the oxygen octahedra occurs in response to 
the confinement electric field; this is the same response that is responsible in some materials (including 
in particular SrTiO$_3$) for extremely large bulk dielectric constants\cite{Khalsa,Neville}.
At the same time atomic-like SO splitting interactions hybridize the three $t_{2g}$ orbitals,
which are split in the 2DES by differences in their confinement energies.  
(For $\langle001\rangle$ 2DESs (assumed below) $xy$ orbitals, which 
have weak bonding along the $z$-direction, have the lowest 
confinement energy and therefore higher occupancy than $\{yz,zx\}$ bands.)  
We explain below how these two effects combine to produce a Rashba interaction.

\begin{figure}[H]
\includegraphics[width=8.6cm]{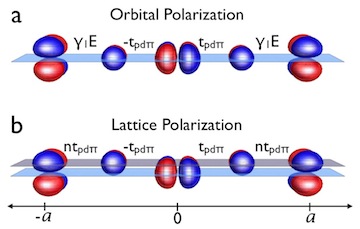} 
\caption{{\bf Bonding network along the y-axis with an electric field. a,} Orbital polarization.  Bonding between $zx$ and $xy$ on neighboring metal atoms through $p_x$ orbitals.  The positive and negative lobes of the orbital functions are represented in blue and red, respectively. {\bf b,} Lattice Polarization.  Displacement of the metal (light blue plane) and oxygen (light purple plane) sublattices in an electric field. 
} 
\label{BondingNetwork}
\end{figure}

The Rashba interaction couples an orbital axial vector that is odd under 
$z \to -z$ to spin, and must therefore arise from hopping process that 
are odd under inversion in the $x-y$ plane.  We therefore begin by 
considering a single plane (see Figure \ref{CrystalStructure}c) of metal atoms 
and identify the relevant process by using 
a tight-binding model\cite{SK} for $p-d$ hybridization,
assigning a hopping amplitude $t_{pd}$ to the process discussed above.  Because of the difference 
in parity between $p$ and $d$ orbitals, $t_{pd}$ changes 
sign when the hopping direction changes  
(Figure \ref{BondingNetwork}).  To leading order in $t_{pd}$ virtual hopping via 
oxygen sites, the Hamiltonian is diagonal in the $t_{2g}$-space 
with eigenenergies:
\bea
\epsilon_{yz} &=& 4 t_1- 2 t_1 \cos(k_y a) \\
\epsilon_{zx} &=& 4 t_1 - 2 t_1 \cos(k_x a) \\ 
\epsilon_{xy} &=& 4 t_2  -2 t_2 cos(k_x a) -2 t_2 cos(k_y a) \\ \nonumber 
\eea 
Here $t_{1,2}=t_{pd}^2  / \Delta_{pd}$ where $\Delta_{pd}$ is the splitting between the oxygen $p$ and metal $t_{2g}$ 
energy levels and the subscripts acknowledge a symmetry allowed difference, ignored below, between $xy$ and $\{yz,zx\}$ hopping amplitudes in the planar environment.  Note that the $xy$ band is twice as wide as the $\{yz,zx\}$ bands 
and lower in energy at the 2D $\Gamma$ point.  Level repulsion from apical oxygens contributes $2t_1$ 
to $\epsilon_{yz,zx}$.  Because effective metal-to-metal hopping amplitudes in this model 
are independent of hopping direction, they do not produce Rashba splitting even when combined with on-site SO terms.    

%Inversion symmetry in the crystal requires that the energy of the system is unchanged by taking $k$ to $Ðk$, while the spin is unchanged.  In the absence of any magnetic fields, the crystal will also have time-reversal symmetry.  Time-reversal symmetry requires that the energy is unchanged is when $k$ to $Ðk$, and the spin is flipped.  Due to both of these symmetries all energy band in the crystal will have at least a two-fold degeneracy (KramerÕs degeneracy)\cite{Kittel}.  By breaking the inversion symmetry in the bulk the KramerÕs degeneracy can be lifted - but this is usually a small effect\cite{Dresselhaus}.   Because the $k=0$ point is still protected by time-reversal symmetry, it remains degenerate.  Near the surface of a crystal, or in an asymmetric confining potential, the inversion symmetry is also lifted Ð leading to a lifting of the KramerÕs degeneracy. In the later case, this splitting depends linearly on electric field and wave-vector\cite{Rashba}. 

Rashba interactions are caused by broken mirror symmetry and in particular by the associated electric field 
$E$ perpendicular to
the 2DES plane.  For $t_{2g}$ 2DESs, this field both polarizes the atomic orbitals and induces a polar 
lattice displacement.  These effects open new covalency channels in the metal-oxygen network.  
In particular there is no hopping in the unperturbed system between a metal $zx$-orbital and an oxygen $x$-orbital 
separated along the $y$-direction.  This is because the $x$-orbital is even and the $zx$-orbital odd 
under reflection in the $xy$-plane passing through the metal-oxygen bond.  When $E \ne 0$, the 
Hamiltonian is no longer invariant under this reflection and the hopping process is allowed.  If, for example,
we think about the perturbation as arising from an additional potential $-eEz$, we can write the induced hopping amplitude
approximately as $E\gamma_1$, where $\gamma_1=   \langle  zx,\vec{R}=0 |-ez  |x,\vec{R}=a/2\hat{y} \rangle $ ({Figure \ref{BondingNetwork}a}). 
At the same time, the electric field will produce forces of opposite sign on metal cations and oxygen anions.  The induced polarization will change the metal oxygen 
bond angle introducing a non-zero $\hat{z}$-component direction cosine 
$n$ in the bond axis direction.  In a 2-center approximation, this change also gives a   
non-zero amplitude $n t_{pd}$ for $zx$ to $x$ hopping along the $y$ direction.
Similar considerations imply an identically induced $yz$ to $y$ hopping amplitude along $x$.  (See Figure \ref{BondingNetwork}b).  
Including these weak effects, which at leading order act only once in the two-step 
metal-oxygen-metal hopping process, we obtain an additional effective metal-to-metal hopping 
amplitude that changes sign with hopping direction and therefore produces a Rashba effect.
The $(yz,zx,xy)$-representation Rashba Hamiltonian is   
\be
\label{HR}
H_{E}^{t_{2g}} = 
\left(
\begin{array}{ccc}
0 & 0 & -2 i t_R \sin(k_x a) \\
 0 & 0& -2 i t_R \sin(k_y a) \\
 2 i t_R \sin(k_x a) & 2 i t_R \sin(k_y a) & 0 \\ 
\end{array}
\right), 
\ee 

\noindent
where the Rashba interaction strength parameter 
$t_R=(\gamma_1 t_{pd} E) / \Delta_{pd} +( n t_{pd}^2)/ \Delta_{pd}$.  
When combined with the an atomic-like bulk SO interaction\cite{Bistritzer,Khalsa}, described in the Supplementary Information, $H_{E}^{t_{2g}}$ leads to Rashba splitting in the $t_{2g}$ bands. 
We remark that broken mirror plane symmetry also introduces other covalent bonding channels, but 
these do not contribute to the Rashba effect.  
We note that, a surface metal atom in a 
BO$_2$ terminated perovskite is not octahedrally coordinated.  
This absence of local inversion symmetry and the decrease in level repulsion with neighboring oxygen atoms mixes $e_g$ and $t_{2g}$ orbitals at the surface.  
When this mixing is strong a more elaborate theory of Rashba SO coupling 
is required.

In general $t_{2g}$ 2DESs will be spread over many coupled metal layers, and the Rashba Hamiltonian
$H_{E}^{t_{2g}}$ will act within each layer with a layer-dependent coupling constant 
$t_{R}$.  For the extreme case of a single-layer $t_{2g}$ 2DES, the ${xy}$- band
will be pulled below the $\{yz,zx\}$-bands by differential confinement effects.
In this case we can derive a simple effective Rashba Hamiltonian which acts 
within the $xy$ subspace.  To do so, we define $\delta$ as the energy scale which 
splits the $xy$ and $\{yz,zx\}$ bands at the $\Gamma$ point.
Allowing virtual transitions to the $\{yz,zx\}$ manifold due to orbital/lattice polarization ($H_E$), and bulk SO effects ($H_{SO}$), we find the part of the Hamiltonian linear in electric field is given at small $k$ by,

\be
\label{HRxy}
H_{R}^{xy} = \epsilon_{xy}(\vec{k})  - \alpha \,  \vec{\sigma} \cdot (\vec{k} \times \hat{z})
\ee

\noindent
where $\alpha = 4 \Delta_{SO} t_R a/(3\delta)$.  

\begin{figure}[H]
\includegraphics[width=8.6cm]{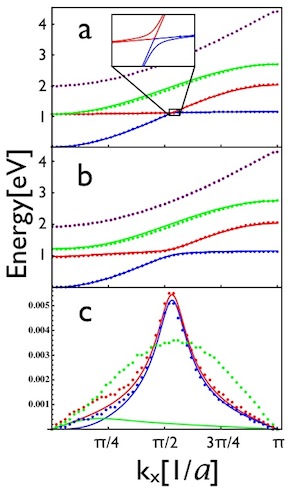} 
\caption{{\bf Orbital polarization changes to $t_{2g}$ bandstructure. a,}  $t_{2g}$ bandstructure in the absence of an electric field and SO coupling.  The results of our model are shown as solid lines while the simulation is shown as dotted lines.  The band order (at the $\Gamma$ point) is $xy$ (blue), $yz$ (red), $zx$ (green), and $x^2-y^2$ (purple).  The inset shows the onset of an avoided crossing in the presence of an electric field. {\bf b,} $t_{2g}$ band structure with SO coupling and orbital polarization.  {\bf c,} Comparison of the orbital polarization part of the Rashba splitting in the $t_{2g}$ space.
} 
\label{OrbitalBands}
\end{figure}

To support our theory of the Rashba effect we have carried out an \emph{ab initio} study of a
typical $t_{2g}$ 2DES.  
To simplify the comparison we examined the case of a single $\langle001\rangle$ BO$_2$ plane and
studied the influences of $z$-direction external electric fields and oxygen-metal sublattice relative displacements  
separately.  
Because we expect Rashba splitting to be proportional to $\Delta_{SO}$ we use the $5d$ 
transition metal Halfnium (Hf) as the B atom.  To minimize the mixing between
$t_{2g}$ and $e_g$ bands apical oxygen atoms have been included in our study 
to maintain octahedral coordination of the metal sites
and maximize crystal field splitting.  
Figure \ref{OrbitalBands}a shows the band structure of a Hf perovskite plane with ideal atomic positions
in the absence of an applied external electric field when spin-orbit interactions are neglected.  
At the zone center, the $xy$ band has a lower energy than the $\{yz,zx\}$ bands as expected in $t_{2g}$ 2DESs.
The strength of the Rashba hopping processes can be read off the band structure by 
identifying the avoided crossing which occurs between an $xz$ or $yz$ band 
and the small mass $xy$ band along a large mass direction in momentum space when an electric field is present.  
We find that even for an extremely large electric field, $0.1eV/$\AA \ , the level repulsion (Eqn. \ref{HR}) at the crossing 
is very small.  Figure \ref{OrbitalBands}b shows the band structure changes when SO coupling is included.  
Note the expected confinement-induced $t_{2g}$ manifold degeneracy lifting at the $\Gamma$ point.
On the scale of this figure the Rashba splitting is too small to be visible.  
Figure \ref{OrbitalBands}c plots the spin splitting, which is largest near the band crossing 
and reaches a maximum value of $\sim 5meV$,  as a function of $k$, and compares it with the splitting predicted by our 
theory when the value of $t_{R}$ is fit to the {\em ab initio} bands calculated in the absence of spin-orbit coupling (see Supplementary Table I for tight-binding fitting parameters).

\begin{figure}[H]
\includegraphics[width=8.6cm]{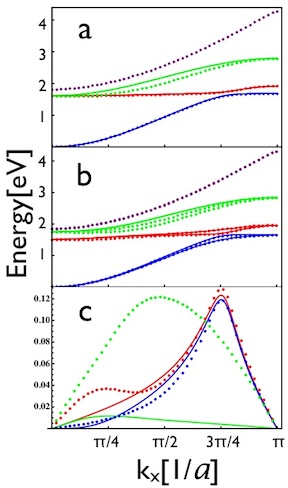} 
\caption{{\bf Lattice polarization changes to $t_{2g}$ bandstructure. a,}  $t_{2g}$ bandstructure in with lattice displacement and no SO coupling.  The results of our model are shown as solid lines while the simulation is shown as dotted lines.  The band order (at the $\Gamma$ point) is $xy$ (blue), $yz$ (red), $zx$ (green), and $x^2-y^2$ (purple).  {\bf b,} $t_{2g}$ band structure with SO coupling and lattice displacement. {\bf  c,} Comparison of the lattice mediated Rashba splitting in the $t_{2g}$ space.
} 
\label{LatticeBands}
\end{figure}

Figure \ref{LatticeBands} reports the corresponding results obtained for the case of 
a polar lattice displacement.  Figure \ref{LatticeBands}a shows the bandstructure for a Hf perovskite plane with a displacement of 0.2 \AA. 
%%
%% Byounghak
The Hf atom displacement was chosen to emulate the interface atomic configuration. 
According to previous first principles calculation in LaTiO$_3$/SrTiO$_3$ interface\cite{Martin}, Ti atoms at the interface are displaced out of the TiO$_2$ plane by $\approx$ 4\% of the lattice constant.
For illustrative purposes we 
report results for a similar Hf atom displacement of $\sim 5\%$ of a lattice constant.

%{\bf(I did calculation of 4\% displacement. The difference is the eg bands that are higher and further %away from t2g bands, compared to 5\%, but otherwise the same.)
% } 
%%
%%
The $yz,xz$-$xy$ avoided crossing is now easily visible.  
After SO coupling is included, a clear Rashba splitting is visible (Figure \ref{LatticeBands}b,c) that 
is an order of magnitude larger than for the orbital polarization case.  
Note that our $t_{2g}$ only model 
underestimates the $zx$ band splitting in both cases, particularly close to the 
band center.  We ascribe this to the proximity of the lower $e_{g}$ level that 
is visible in the band plots and neglected in our theory.

Our theory of Rashba interactions in $t_{2g}$ 2DESs is consistent with experimental 
evidence for strong spin-orbit interactions at interfaces between polar and 
non-polar perovskites\cite{Caviglia,Berakdar,Triscone}, for example the SrTiO$_3$/LaAlO$_3$ interface, and in 
surface 2DESs induced by the very strong electric fields
applied by ionic liquid gates.  It also suggests that Rashba interactions will tend to 
be stronger in materials which are more easily polarized.  In this sense SrTiO$_3$ has a potential for relatively strong spin-orbit interactions even 
though Ti is a 3D material.  $t_{2g}$ 2DESs with strong lattice polarizabilities
may\cite{Khalsa} contain both weakly-confined orbitals 
responsible for superconductivity and strongly-confined orbitals 
responsibility for magnetism\cite{PALee}.  If so, 
our theory implies that magneto-transport properties in these materials
will be strongly sensitive to local lattice polarization at the 
surface or interface.  
%By comparing with suggested band parameters for SrTiO$_3$ 2DESs \cite{Khalsa} and a polar displacement\cite{Martin} of $4\%$ we find that the Rashba interaction strength parameter near the SrTiO$_3$/LaAlO$_3$ interface may be as large as 9meV.  
We believe that this work provides a starting point
for the interpretation of heretofore unexplained magneto-transport phenomena \cite{Dagan,Ilani}.

{\bf Methods:} The first principles calculations were based on the density functional theory and carried out using the Vienna {\it Ab Initio} Simulation Package\cite{vasp}.
We used projector-augmented wave pseudopotentials and the generalized gradient approximation exchange-correlation functional of Perdew, Burke and Ernzerhof\cite{pbe}. 
The supercell contained a three atom HfO$_2$ layer with two oxygen atoms located directly above and below the Hf atom. The molecular layers were separated by 20 \AA~ vacuum. 
The plane-wave energy cutoff was set to 500 eV.
We employed a $8 \times 8 \times 2$ $k$-point sampling to achieve electronic convergence.

\pagebreak

\noindent
{\bf Acknowledgements:}
This work has been supported by the Welch Foundation under grant TBF1473 and by the National Science Foundation under grant DMR-1122603. BL was supported by US AFOSR through contract FA9950-10-1-0133.

\noindent
{\bf Competing Financial Interests Statement}
The authors declare no competing financial interests. 

\noindent
{\bf Author Contributions}
All authors contributed to developing the theory and preparing the manuscript. G.K and B.L. contributed to the DFT calculations.

\newpage

\clearpage

\section*{Supplementary Information} 

The explicit form of the atomic-like SO coupling Hamiltonian\cite{Khalsa2,Bistritzer2} mentioned in the main text is given in the $(yz\uparrow, zx\uparrow, xy\uparrow, yz\downarrow, zx\downarrow, xy\downarrow)$ representation by,

\end{spacing}
\begin{spacing}{1}

\be
\tag{S1}
\label{SO}
H^{SO}  = \frac{\Delta_{SO}}{3}
\left(
\begin{array}{cccccc}
0 & i & 0 & 0 & 0 & -1 \\
-i & 0 & 0 & 0 & 0 & i \\
0 & 0 & 0 & 1 & -i & 0 \\
0 & 0 & 1 & 0 & -i & 0 \\
0 & 0 & i & i & 0 & 0 \\
-1 & -i & 0 & 0 & 0 & 0 \\
\end{array}
\right).
\ee 

\end{spacing}
\vspace{8mm}
\begin{spacing}{2}

Breaking the mirror symmetry of the 2DES allows additional covalent bonding terms - not present in the unperturbed cubic system.  Those described in the main text are associated with the Rashba effect, but others new processes are allowed.  These are summarized in the Table II.  In the unpolarized system, the $p_x$ orbital is not involved in the bonding along the $x$ direction.  For this reason $\gamma_2$ processes, which couples $zx$ and $p_x$ along the $x$ direction, can only contribute to second order in $E$.  The corresponding statement is also true for lattice displacements.  

$\gamma_3$ increases or decreases the bonding strength with the apical oxygens.  If both apical oxygen atoms are present, it can only contribute a term in the Hamiltonian that is quadratic in electric field.  At the surface of a BO$_2$ terminated perovskite, there is a missing apical oxygen. Therefor, within the $t_{2g}$ space and to first order in the electric field, the following term must be added to the Hamiltonian of the surface layer:

\be
\tag{S2}
\label{HS}
H_{S} = 
\left(
\begin{array}{ccc}
\Delta_S & 0 & 0 \\
 0 & \Delta_S & 0 \\
0 & 0 & 0 \\
\end{array}
\right).
\ee 

In the above equation, $\Delta_S = - E \gamma_3 t_{pd\pi}/\Delta_{pd}$ and may be positive or negative.  Because there is no periodicity in the $z$- direction, this term is independent of $k$.  In addition, the lack of octahedral coordination at the surface can lead to significant $t_{2g}/e_g $ hybridization.

In addition to the bonding changes to the cubic system mentioned above, if the structural distortions are more complicated (e.g. there is an in-plane twisting of the oxygen octahedra) some $\sigma$ bonding between the oxygen $p$ and metal $t_{2g}$ orbitals will also be allowed.  Provided the polar displacement is still present, these $\sigma$ bonding channels can also contribute to the Rashba effect.

\begin{table}
\begin{tabular}{| l | l | l |}

\hline
\multicolumn{3}{| c |}
{Tight-Binding Model Parameters} \\
\hline

\multirow{1}{*}
Lattice Constant & $a$ & 4.05 \AA \\ 
SO splitting & $\Delta_{SO}$ & 0.340 eV \\
\hline

\multirow{4}{*}
{Orbital Polarization} & $t_1$ & 0.41 eV \\
 & $t_2$ & 0.51 eV \\
 & $t_R$ & 0.0014 eV \\
 & $t'$ & 0.02 eV \\
  \hline
  
\multirow{4}{*}
{Lattice Polarization} & $t_1$ & 0.30 eV \\
 & $t_2$ & 0.48 eV \\
 & $t_R$ & 0.045 eV \\
 & $t'$ & 0.02 eV \\
 \hline

\end{tabular}
\label{Table1}
\caption{Tight-binding parameters used in fitting \emph{ab initio} results for HfO$_2$ plane.}
\end{table}

\begin{table}
\begin{tabular}{| l | l | l |}

\hline
\multicolumn{3}{| c |}
{Tight-Binding Matrix Elements} \\
\hline

\multirow{1}{*}
Lattice Polarization Effects & $\langle xy ,\vec{R}=0|\Delta U|\{x,y,z\} ,\vec{R} = \pm \frac{a}{2} \hat{x}+n \frac{a}{2} \hat{z} \rangle$ & $\{0,\pm t_{pd\pi},0\}$  \\ 
 & $\langle xy ,\vec{R}=0|\Delta U|\{x,y,z\} ,\vec{R} = \pm \frac{a}{2} \hat{y} +n \frac{a}{2} \hat{z} \rangle$ & $\{\pm t_{pd\pi},0,0\}$ \\
 & $\langle yz ,\vec{R}=0|\Delta U|\{x,y,z\} ,\vec{R} = \pm \frac{a}{2} \hat{x} +n \frac{a}{2} \hat{z} \rangle$ & $\{0,n t_{pd\pi},0\}$ \\
 & $\langle yz ,\vec{R}=0|\Delta U|\{x,y,z\} ,\vec{R} = \pm \frac{a}{2} \hat{y} +n \frac{a}{2} \hat{z} \rangle $ & $\{0,n t_{pd\pi},0\}$ \\
 
\hline

\multirow{1}{*}
{Orbital Polarization Effects} & $\langle xy ,\vec{R}=0|-ez|\{x,y,z\} ,\vec{R}=\pm \frac{a}{2} \hat{x} \rangle
$ & $\{0,0,0\}$ \\
 & $\langle xy ,\vec{R}=0|-ez|\{x,y,z\} ,\vec{R}=\pm \frac{a}{2} \hat{y} \rangle
$ & $\{0,0,0\}$ \\
 & $\langle yz ,\vec{R}=0|-ez|\{x,y,z\} ,\vec{R}=\pm \frac{a}{2} \hat{x} \rangle
$ & $ \{0,\gamma_1,0\}$ \\
 & $\langle yz ,\vec{R}=0|-ez|\{x,y,z\} ,\vec{R}=\pm \frac{a}{2} \hat{y} \rangle
$ & $ \{0,\gamma_2,0\}$ \\  
  \hline
  
\multirow{1}{*}
{Apical Oxygen Atoms} & $\langle xy ,\vec{R}=0|-ez|\{x,y,z\} ,\vec{R}=\pm \frac{a}{2} \hat{z} \rangle
$ & $ \{0,0,0\}$ \\
 & $\langle yz ,\vec{R}=0|-ez|\{x,y,z\} ,\vec{R}=\pm \frac{a}{2} \hat{z} \rangle
$ &$\{0,\pm \gamma_3,0\}$\\
 \hline

\end{tabular}
\label{Table2}
\caption{Tight-binding matrix elements for metal $t_{2g}$ and oxygen p-orbitals.  The matrix elements for the $zx$ orbitals can be derived from the $yz$ entries, by symmetry.}
\end{table}

\end{spacing}


\begin{thebibliography}{99}

\bibitem{Kittel} Kittel, C. {\em Quantum Theory of Solids} (Wiley, New York, 1963). 
%(*symmetry argument for degeneracy Ð cited in Winkler*)

\bibitem{Dresselhaus} Dresselhaus, G. Spin-orbit coupling effects in zinc blende structures, Phys. Rev. {\bf 100}, 580Ð586 (1955). 
%(*bulk inversion asymmetry Ð cited in Winkler*)

\bibitem{Rashba} Bychkov, \& Y.A., Rashba, J. Oscillatory effects and the magnetic susceptibility of carriers in inversion layers.  J. Phys. C: Solid State {\bf 17}, 6039-6045 (1984). 
%(*surface inversion asymmetry Ð cited in Winkler*)

\bibitem{Winkler} Winkler, R.  Spin-Orbit Coupling Effects in Two-Dimensional Electron and Hole Systems, Springer Tracts in Modern Physics {\bf 191}, (Springer, Berlin, 2003).
% (*general reading*)

\bibitem{Hwang} Ohtomo, A., \& Hwang, H.Y.  A high-mobility electron gas at the LaAlO3/SrTiO3 heterointerface Nature {\bf 427}, 423 (2004). 
% (*general oxide 2DEG*)

\bibitem{Mannhart} Thiel, S., Hammerl, G., Schmehl, A.,  Schneider, C.W., \& Mannhart, J.  Tunable Quasi-Two-Dimensional Electron Gases in Oxide Heterostructures.  Science {\bf 313}, 1942 (2006).
% (*general oxide 2DEG*)

\bibitem{Ueno2008} Ueno, K., Nakamura, S., Shimotani, H. \& Ohtomo, A. Electric-field-induced superconductivity in an insulator. Nat. Mater. {\bf 7}, 855 (2008). 
% (*SrTiO3 2DEG electric field*)

\bibitem{Ueno2011} Ueno, K., Nakamura, S. \& Shimotani, H. Discovery of superconductivity in KTaO3 by electrostatic carrier doping. Nature Nanotechnol. {\bf 6}, 408 (2011). 
% (*KTaO3 2DEG electric field*)

\bibitem{OxideInterfaces} 
Mannhart, J. \& Schlom, D.G.  Oxide Interfaces Ð An Opportunity for Electronics.  Science {\bf 327}, 5973 (2010).  
%(* oxide interfaces general motivation *)

\bibitem{Goodenough} Goodenough, J.B. {\em Localized to Itinerant Electronic Transition in
Perovskite Oxides} (Springer, Berlin, 1996).

\bibitem{Mattheiss} Mattheiss, L.F.  Energy Bands for KNiF3, SrTiO3, KMoO3, and KTaO3.  Phys. Rev. B {\bf 6}, 4718-4740 (1972).
%(* general electronic structure of Perovskites *)

\bibitem{Khalsa} Khalsa, G. \& MacDonald, A.H. Theory of the SrTiO$_3$ surface state two-dimensional electron gas.  Phys. Rev. B {\bf 86}, 125121 (2012).
%(*previous 2DEG modeling*)

\bibitem{Stengel} Stengel, M.  First-Principles Modeling of Electrostatically Doped Perovskite Systems. Phys. Rev. Lett. {\bf 106}, 136803 (2011).  
%(* previous 2DEG modeling *)

\bibitem{Bistritzer} Bistritzer, R., Khalsa, G, \& MacDonald, A.H. Electronic structure of doped d0 perovskite semiconductors.  Phys. Rev. B {\bf 83}, 115114 (2011).
% SO Hamiltonian

\bibitem{Neville} Neville, R.C., Hoeneisen, B., \& Mead C.A.  Permittivity of Strontium Titanate.  J. Appl. Phys. {\bf 43}, 2124 (1972).  
%(*SrTiO3 permittivity study*)

\bibitem{SK} Slater, J.C., \& Koster, G.F. Simplified LCAO Method for the Periodic Potential Problem.  Phys. Rev. {\bf 94}, 1498Ð1524 (1954).  
%(*tight binding*)

\bibitem{Martin} Popovi\'c, Z.S., Satpathy, S., \& Martin, R.M. Origin of the two-dimensional electron gas carrier density at the LaAlO$_3$ on SrTiO$_3$ interface. Phys. Rev. Lett. {\bf 101}, 256801 (2008).
% Richard Martin LAO/STO DFT

\bibitem{Caviglia} Caviglia, A.D. Tunable Rashba Spin-Orbit Interaction at Oxide Interfaces.  Phys. Rev. Lett. {\bf 104}, 126803 (2010). 
%(*Rashba LAO/STO*)

\bibitem{Berakdar} Jia, C. \& Berakdar, J. Magnetotransport and spin dynamics in an electron gas formed at oxide interfaces. Phys. Rev. B {\bf 83}, 045309 (2011).
% Rashba

\bibitem{Triscone} Fete, A., Gariglio, S., Caviglia, A.D., Triscone, J.M., \& Gabay, M. Rashba induced magnetoconductance oscillations in the LaAlO3-SrTiO3 heterostructure.  Phys. Rev. B {\bf86}, 201105(R) (2012). 
%(*Rashba LAO/STO*)

\bibitem{PALee} Michaeli, K., Potter, A.C., \& Lee, P.A. 
Superconducting and Ferromagnetic Phases in SrTiO3/LaAlO3 Oxide Interface Structures: Possibility of Finite Momentum Pairing. Phys. Rev. Lett. {\bf 108}, 117003 (2012).

\bibitem{Dagan}  Flekser, E. et al. Magnetotransport effects in polar versus non-polar SrTiO$_{3}$ based heterostructures. Phys. Rev. B {\bf 86}, 121104 (2012).
%magneto-transport

\bibitem{Ilani} Joshua, A., Ruhman, J., Pecker, S., Altman, E., \& Ilani, S. Unconventional Phase Diagram of Two-Dimensional Electrons at the LaAlO3/SrTiO3 Interface.  arXiv:1207.7220. 
%Magneto-transport

\bibitem{vasp} Kresse, G.  \& FurthmŸller, J.  
Efficiency of ab-initio total energy calculations for metals and semiconductors using a plane-wave basis set. 
Comput. Mater. Sci. {\bf 6}, 15 (1996).

\bibitem{pbe} Perdew, J. P., Burke K., \& Ernzerhof, M. 
Generalized Gradient Approximation Made Simple.
Phys. Rev. Lett. {\bf 77}, 3865 (1996).

%\bibitem{Uemura} Ohkawa, F.J. \& Uemura, Y. Quantized Surface States of a Narrow-Gap Semiconductor J. Phys. Soc. Jpn. 
%{\bf 37}, 1325 (1974).
%(*surface inversion asymmetry Ð cited in Winkler*) 

%\bibitem{Demkov} Lee, Jaekwang, \& Demkov, Alexander A. Charge origin and localization at the n-type SrTiO3/LaAlO3 
%interface, Phys. Rev. B {\bf 78}, 193104 (2008). 

\end{thebibliography}

\newpage

\begin{thebibliography}{99}

\bibitem{Khalsa2} Khalsa, G. \& MacDonald, A.H. Theory of the SrTiO$_3$ surface state two-dimensional electron gas.  Phys. Rev. B {\bf 86}, 125121 (2012).
%(*previous 2DEG modeling*)

\bibitem{Bistritzer2} Bistritzer, R., Khalsa, G, \& MacDonald, A.H. Electronic structure of doped d0 perovskite semiconductors.  Phys. Rev. B {\bf 83}, 115114 (2011).
% SO Hamiltonian

\end{thebibliography}
\end{document}